\documentclass[preprint,showpacs,preprintnumbers,amsmath,amssymb]{revtex4}

\usepackage{graphicx}
\usepackage{dcolumn}
\usepackage{bm}
\usepackage{braket}

\newtheorem{lem}{Lemma}

\begin{document}


\title{Quantum Entanglement Capacity with Classical Feedback}

\author{Alan W. Leung}
 \email{leung@math.mit.edu}
\affiliation{%
Department of Mathematics, Massachusetts Institute of
Technology,\\77 Massachusetts Avenue, Cambridge, MA 02139, USA
}%

\date{\today}

\begin{abstract}
For any quantum discrete memoryless channel, we define a quantity
called quantum entanglement capacity with classical feedback
($E_B$), and we show that this quantity lies between two other
well-studied quantities. These two quantities - namely the quantum
capacity assisted by two-way classical communication ($Q_2$) and
the quantum capacity with classical feedback ($Q_B$) - are widely
conjectured to be different: there exists quantum discrete
memoryless channel for which $Q_2>Q_B$. We then present a general
scheme to convert any quantum error-correcting codes into adaptive
protocols for this newly-defined quantity of the quantum
depolarizing channel, and illustrate with Cat (repetition) code
and Shor code. We contrast the present notion with entanglement
purification protocols by showing that whilst the Leung-Shor
protocol can be applied directly, recurrence methods need to be
supplemented with other techniques but at the same time offer a
way to improve the aforementioned Cat code. For the quantum
depolarizing channel, we prove a formula that gives lower bounds
on the quantum capacity with classical feedback from any $E_B$
protocols. We then apply this formula to the $E_B$ protocols that
we discuss to obtain new lower bounds on the quantum capacity with
classical feedback of the quantum depolarizing channel.
\end{abstract}

\maketitle

\section{Introduction}\label{sec:intro}

Quantum information theory\cite{N7,B5} studies transmission and
manipulation of information in systems that must be treated
quantum mechanically, and it is markedly different from classical
information theory\cite{B2,N8} in which the capacity of a
classical discrete memoryless channel is uniquely given by a
single numerical value representing the amount of information that
can be transmitted asymptotically without error per channel use.
Moreover, this value is unaffected by the use of classical
feedback. However, for quantum discrete memoryless channels,
capacities are affected by side classical communication and shared
entanglements\cite{1,6,7}. In addition, we can use a quantum
channel to transmit either classical or quantum information and
therefore we can define, for every quantum discrete memoryless
channel, various capacities: $C$, unassisted classical capacity;
$C_B$, classical capacity assisted by classical feedback; $C_2$,
classical capacity assisted by independent classical information;
$C_E$, entanglement-assisted classical capacity; $Q$, unassisted
quantum capacity; $Q_B$, quantum capacity assisted by classical
feedback; $Q_2$, quantum capacity assisted by independent
classical information; and finally $Q_E$, entanglement-assisted
quantum capacity.

So far, some progress has been made to compute the capacities for
specific channels\cite{17,6,13}. However, search for a general
formula only succeeded in a few cases\cite{7,39,2,3,8}, and
progress in this direction has been hindered by the additivity
conjecture\cite{28,29,41,9}. While we are far from obtaining a
formula for all these capacities, a natural question to ask is
whether we can relate these capacities. Some relations such as
$C\geq Q$ are trivial but others can be hard. Some capacities are
even incomparable, i.e. depending on the channel, either one may
be greater than the other. For the comparable capacities, we also
want to show whether the inequalities are strict or saturable. Our
present knowledge of these relations is summarized in \cite{1, 5}.

One of the conjectural relations is $Q_2>Q_B$, that there exist
quantum channels whose quantum capacity assisted by two-way
classical communication exceeds their quantum capacity assisted by
classical feedback. While we cannot prove the conjecture, the aim
of this work is to define, for any quantum discrete memoryless
channel, a quantity called quantum entanglement capacity with
classical feedback ($E_B$). We show that this capacity lies
between $Q_B$ and $Q_2$, and it has two different well-defined
operational meanings. For the quantum depolarizing channel, we
demonstrate a general scheme to convert quantum error-correcting
codes (QECC) into $E_B$ protocols, and these in turn imply new
lower bounds on the quantum capacity with classical feedback
($Q_B$).

This work is also closely related to entanglement purification
protocols (EPP)\cite{35,N18,LS1,51,N17}, procedures by which two
parties can extract pure-state entanglement out of some shared
mixed entangled states. For example,

\begin{eqnarray}
00:&\ket{\Phi^{+}}=\frac{1}{\sqrt{2}}(\ket{\uparrow\uparrow}+\ket{\downarrow\downarrow})\nonumber\\
01:&\ket{\Psi^{+}}=\frac{1}{\sqrt{2}}(\ket{\uparrow\downarrow}+\ket{\downarrow\uparrow})\nonumber\\
10:&\ket{\Phi^{-}}=\frac{1}{\sqrt{2}}(\ket{\uparrow\uparrow}-\ket{\downarrow\downarrow})\nonumber\\
11:&\ket{\Psi^{-}}=\frac{1}{\sqrt{2}}(\ket{\uparrow\downarrow}-\ket{\downarrow\uparrow})\label{eqn:
bell states}
\end{eqnarray}

\noindent are the so-called Bell basis and each of these states is
considered equivalent to an ebit, a basic unit of entanglement in
quantum information theory. At the beginning of these entanglement
purification protocols, two persons Alice and Bob share a large
number of the generalized Werner states\cite{N14}

\begin{eqnarray}
\rho_F = F \ket{\Phi^{+}}\bra{\Phi^{+}}+\frac{1-F}{3}\big(
\ket{\Phi^{-}}\bra{\Phi^{-}} + \ket{\Psi^{+}}\bra{\Psi^{+}} +
\ket{\Psi^{-}}\bra{\Psi^{-}}\big), \label{eqn: werner states}
\end{eqnarray}

\noindent say $\rho_F^{\otimes N}$, and they are allowed to
communicate classically, apply unitary transformations and perform
projective measurements. In the end the quantum states $\Upsilon$
shared by Alice and Bob are to be a close approximation of the
maximally entangled states $(\ket{\Phi^{+}}
\bra{\Phi^{+}})^{\otimes M}$, or more precisely we require the
fidelity between $\Upsilon$ and $(\ket{\Phi^{+}}
\bra{\Phi^{+}})^{\otimes M}$ approaches one as $N$ goes to
infinity. We then define the yield of such protocols to be $M/N$.
Entanglement purification protocols (EPP) are further divided into
1-EPP and 2-EPP according to whether the sender and receiver are
allowed to communicate uni- or bi-directionally.

One of the main reasons why this is considered general is the
equivalence between an entanglement purification protocol on the
Werner state $\rho_F$ and a protocol to faithfully transmit
quantum states through the $\frac{(4F-1)}{3}$-depolarizing channel
established in \cite{35}. A $p$-depolarizing channel is a simple
qubit channel such that a qubit passes through the channel
undisturbed with probability $p$ and outputs as a completely
random qubit with probability $1-p$. Specifically, the yield of a
1-EPP on the Werner state $\rho_F$ is equal to the unassisted
quantum capacity of a $\frac{(4F-1)}{3}$-depolarizing channel
($Q$); and the yield of a 2-EPP on the Werner state $\rho_F$ is
equal to the quantum capacity assisted by two-way classical
communication of a $\frac{(4F-1)}{3}$-depolarizing channel
($Q_2$). The equivalence was proved by noting that the EPR pair
$\ket{\Phi^+}$ becomes the Werner state $\rho_F$ if Alice passes
the second half through the $\frac{(4F-1)}{3}$-depolarizing
channel. The present study of the amount of entanglements Alice
and Bob can share by using the depolarizing channel and classical
feedback is clearly related, and we will exploit the similarities
and differences to obtain new results and ask new questions.

\subsection{Structure of the paper}
In section \ref{sec:previous}, we review some previous
entanglement purification protocols that will be used in this
paper. In section \ref{sec: E_B-between}, we define a new quantity
called quantum entanglement capacity with classical feedback
($E_B$) and this quantity is shown to lie between $Q_B$ and $Q_2$.
We will then give an alternate operational meaning of $E_B$. In
section \ref{sec: AQECC}, we describe how one can turn a QECC into
an $E_B$ protocol and illustrate the idea with Cat (repetition)
code and Shor code. We then connect the present notion to the
modified recurrence method and Leung-Shor method. In section
\ref{sec: new-Q_B}, we compute new lower bounds on $Q_B$ implied
by these $E_B$ protocols. Finally, we conclude with a
characteristic of the threshold of Cat code and other further
research directions.

\subsection{Previous works}\label{sec:previous}

\subsubsection{Universal hashing}

Universal hashing, introduced in \cite{35}, requires only one-way
classical communication and hence is a 1-EPP. The hashing method
works by having Alice and Bob each perform some local unitary
operations on the corresponding members of the shared bipartite
quantum states. They then locally measure some of the pairs to
gain classical information about the identities of the the
remaining unmeasured pairs. It was shown that each measurement can
be made to reveal almost 1 bit of information about the unmeasured
Bell states pairs. Since the information associated with a quantum
state $\rho_F$ is given by its von Neumann entropy $S(\rho_F)$, we
know from typical subspace argument that, with probability
approaching 1 and by measuring $N S(\rho_F)$ pairs, Alice and Bob
can figure out the identities of all pairs including the
unmeasured ones. Once the identities of the Bell states are known,
Alice and Bob can convert them into the standard states $\Phi^{+}$
easily. Therefore this protocol distills a yield of $\Big(N -
NS(\rho_F)\Big)/N=1-S(\rho_F)$.

\subsubsection{The recurrence method and the modified recurrence
method} \label{sec: recurrence-method-1}

\begin{figure}[h!]
\begin{center}
\includegraphics[width=60mm]{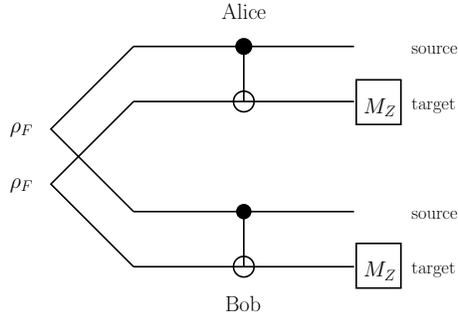}
\caption{\label{fig: 1-recurrence} The recurrence method.}
\end{center}
\end{figure}

The recurrence method\cite{N12,35} is illustrated in figure
\ref{fig: 1-recurrence}. Alice and Bob put the quantum states
$\rho_F^{\otimes N}$ into groups of two and apply XOR operations
to the corresponding members of the quantum states
$\rho_F^{\otimes 2}$, one as the source and one as the target.
They then take projective measurements on the target states along
the z-axis, and compare their measurement results with the side
classical communication channel. If they get identical results,
the source pair ``passed"; otherwise the source pair ``failed".
Alice and Bob then collect all the ``passed" pairs, and iterate
this process until it becomes more beneficial to pass on to the
universal hashing. If we denote the quantum states by
$\rho=p_{00}\ket{\Phi^{+}}\bra{\Phi^{+}}
+p_{01}\ket{\Psi^{+}}\bra{\Psi^{+}} +p_{10}\ket{\Phi^{-}}
\bra{\Phi^{-}}+p_{11}\ket{\Psi^{-}}\bra{\Psi^{-}}$, then this
protocol has the following recurrence relation:

{\setlength\arraycolsep{2pt}
\begin{eqnarray}
&p'_{00}=(p_{00}^2+p_{10}^2)/p_{pass};
& p'_{01}=(p_{01}^2 +p_{11}^2)/p_{pass}; \nonumber\\
&p'_{10}=2p_{01}p_{11}/p_{pass};
&p'_{11}=2p_{00}p_{10}/p_{pass};\label{eqn: recurrence-method1}
\end{eqnarray}}

\noindent and

{\setlength\arraycolsep{2pt}
\begin{eqnarray}
p_{pass}=p_{00}^2+p_{01}^2+p_{10}^2+p_{11}^2+2p_{00}p_{10}+2p_{01}p_{11}.
\label{eqn: recurrence-method2}
\end{eqnarray}}

\noindent This is known as the recurrence method. As mentioned in
\cite{35}, C. Macchiavello has found that if we apply a unilateral
$\pi$ rotation $\sigma_x$ followed by a bilateral $\pi/2$ rotation
$B_x$, faster convergence is achieved and this is known as the
modified recurrence method. Computationally, one has to switch the
$p_{10}$ and $p_{11}$ components after each recurrence.

\subsubsection{The Leung-Shor method} \label{sec: LS}

\begin{figure}[h!]
\begin{center}
\includegraphics[width=60mm]{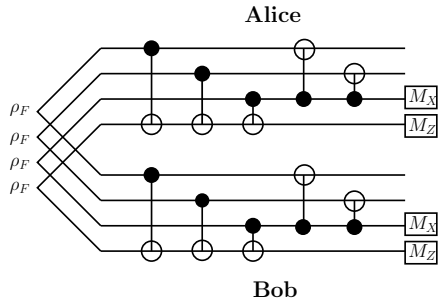}
\caption{\label{fig: 2-LS}The Leung-Shor method.}
\end{center}
\end{figure}

The Leung-Shor method\cite{LS1} is illustrated in figure \ref{fig:
2-LS}. Alice and Bob share the quantum states $\rho_F^{\otimes N}$
and put them into groups of four. They then apply the quantum
circuit shown in figure \ref{fig: 2-LS} and take measurements on
the third and fourth pairs along the x- and z-axis respectively.
Using the side classical communication channel, they can compare
their results with each other. If they get identical results on
both measurements, they keep the first and second pairs and apply
universal hashing\cite{35}. If either of the two results
disagrees, they throw away all four pairs.

The four pairs can be described by an 8-bit binary string as in
(\ref{eqn: bell states}), and since these are mixed states they
are in fact probability distribution over all $256 (=2^8)$
possible 8-bit binary strings. The quantum circuit consists only
of XOR gates and therefore maps the 8-bit binary strings, along
with their underlying probability distribution, bijectively to
themselves. If we let the probability distributions before and
after the quantum gates to be $P(a_1 a_2 b_1 b_2 c_1 c_2 d_1 d_2)$
and $P'(a_1 a_2 b_1 b_2 c_1 c_2 d_1 d_2)$ respectively, then the
yield of this method is:

\begin{equation}
\frac{p_{pass}}{2}\bigg(1-\frac{H(Q(a_1a_2b_1b_2))}{2}\bigg)
\label{eqn: closed-form1}
\end{equation}

\noindent where $p_{pass}=\sum_{a_1,a_2,b_1,b_2,
c_2,d_1\in\{0,1\}}P'(a_1 a_2 b_1 b_2 0c_2d_10)$ is the ``pass"
probability, $Q(a_1a_2b_1b_2)=\sum_{c_2,d_1\in\{0,1\}}P'(a_1 a_2
b_1 b_2 0c_2d_10)/p_{pass}$ is the post-measurement probability
distribution and $H(Q(a_1a_2b_1b_2))$ is the Shannon entropy
function.

\section{A quantity that lies between $Q_B$ and $Q_2$} \label{sec:
E_B-between} In this section, we define, for any quantum discrete
memoryless channel, a quantity called quantum entanglement
capacity with classical feedback $E_B$. We will show that this
quantity is less than the quantum capacity with two-way classical
communication $Q_2$ and is greater than the quantum capacity with
classical feedback $Q_B$.

\subsection{Definition of $E_B$} \label{sec: E_B-definition}
Quantum entanglement capacity with classical feedback of a QDMC
can be loosely described as the maximal asymptotic rate at which
the sender Alice can share the entangled state
$\ket{\Phi^{+}}\in\mathcal{H}_2^{\otimes 2}$ with the receiver Bob
with the assistance of a classical feedback channel. Precisely,
let the QDMC be described by

\begin{eqnarray}
\mathcal{N}: \mathcal{B}(\mathcal{H}_{d_1}) \longrightarrow
\mathcal{B}(\mathcal{H}_{d_2})\nonumber \\
\rho \mapsto \rho' = \sum_{i} E_i \rho E_i^{\dagger}, \nonumber
\end{eqnarray}

\noindent where $\sum_j E_j^{\dagger}E_j=I$ and $\{E_i\}$ is a set
of linear operators which map the input Hilbert space
$\mathcal{H}_{d_1}$ to the output Hilbert space
$\mathcal{H}_{d_2}$. Then in the first round of any $E_B$
protocols, Alice prepares a quantum state
$\alpha_1=\ket{\Upsilon}\bra{\Upsilon}\in \mathcal{B}(
\mathcal{H}_{d_1}^{\otimes N} \otimes\mathcal{H}_a)$, where
$\mathcal{H}_a$ is the Hilbert space representing the ancilla
system in her laboratory and she sends the first part of the
quantum state to Bob via the quantum channel $\mathcal{N}$:

\begin{eqnarray}
\mathcal{N}: \mathcal{B}(\mathcal{H}_{d_1}) &\longrightarrow&
\mathcal{B}(\mathcal{H}_{d_2})\nonumber \\
\rho_1=tr_{(d_1^{N-1}\times a)}(\ket{\Upsilon}\bra{\Upsilon})
&\mapsto& \rho_1' = \sum_{i} E_i \rho_1 E_i^{\dagger}. \nonumber
\end{eqnarray}

\noindent After sending $\rho_1$, Alice's quantum system is
described by $\alpha_1' = tr_{d_1}(\alpha_1)\in \mathcal{B}(
\mathcal{H}_{d_1}^{\otimes(N-1)}\otimes\mathcal{H}_a)$. On the
other hand, Bob is now in possession of the quantum state
$\rho_1'$ he just received from Alice as well as the ancilla
system in his laboratory, and therefore his quantum system can be
described by $\beta_1'= \rho_1'\otimes\beta_1=\rho_1'\otimes
\ket{0}\bra{0}^{\otimes \log_2 b} \in
\mathcal{B}(\mathcal{H}_{d_2}\otimes \mathcal{H}_b)$. Next Bob
performs local quantum operation on his quantum system:

\begin{eqnarray}
\mathbf{B}: \mathcal{B}(\mathcal{H}_{d_2}\otimes \mathcal{H}_b)
&\longrightarrow& \mathcal{B}(\mathcal{H}_{d_2}\otimes \mathcal{H}_b) \nonumber\\
\beta_1' &\mapsto& \beta_1'' = \sum_{i} B_i \beta_1'
B_i^{\dagger}\nonumber
\end{eqnarray}

\noindent where $\sum_i B_i^{\dagger}B_i=I$. Bob then uses the
feedback channel to send classical information to Alice. Note that
if Bob's operation comprised quantum measurements, this classical
information could include the measurement results$(i)$. Upon
learning the classical information sent by Bob, Alice's quantum
system transforms from $\alpha_1'$ to $\alpha_{1,(i)}'$ and she
performs operation on her quantum system:

\begin{eqnarray}
\mathbf{A_{(i)}}: \mathcal{B}(\mathcal{H}_{d_1}^{\otimes(N-1)}
\otimes\mathcal{H}_a) &\longrightarrow& \mathcal{B}(
\mathcal{H}_{d_1}^{\otimes(N-1)}\otimes\mathcal{H}_a) \nonumber\\
\alpha_{1,(i)}' &\mapsto& \alpha_{1,(i)}'' = \sum_{j} A_{j,(i)}
\alpha_{1,(i)}' A_{j,(i)}^{\dagger}.\nonumber
\end{eqnarray}

\noindent Note that both the quantum system $\alpha_{1,(i)}'$ and
Alice's operation $\mathbf{A_{(i)}}$ are dependent on the
classical information(i) she received from Bob. This is the end of
the first round of any general $E_B$ protocols and can be
summarized as:

\begin{eqnarray}
\mathbf{LOCC_{A\leftarrow B}}^{(1)}\circ\mathbf{N}^{(1)}:
\mathcal{B}(\mathcal{H}_{d_1}^{\otimes
N}\otimes\mathcal{H}_a\otimes\mathcal{H}_b) &\longrightarrow&
\mathcal{B}(\mathcal{H}_{d_1}^{\otimes
(N-1)}\otimes\mathcal{H}_a\otimes\mathcal{H}_{d_2}\otimes\mathcal{H}_b)
\nonumber\\
\omega_1 &\mapsto& \omega_2. \nonumber
\end{eqnarray}

\begin{figure}
\begin{center}
\includegraphics[height=220mm]{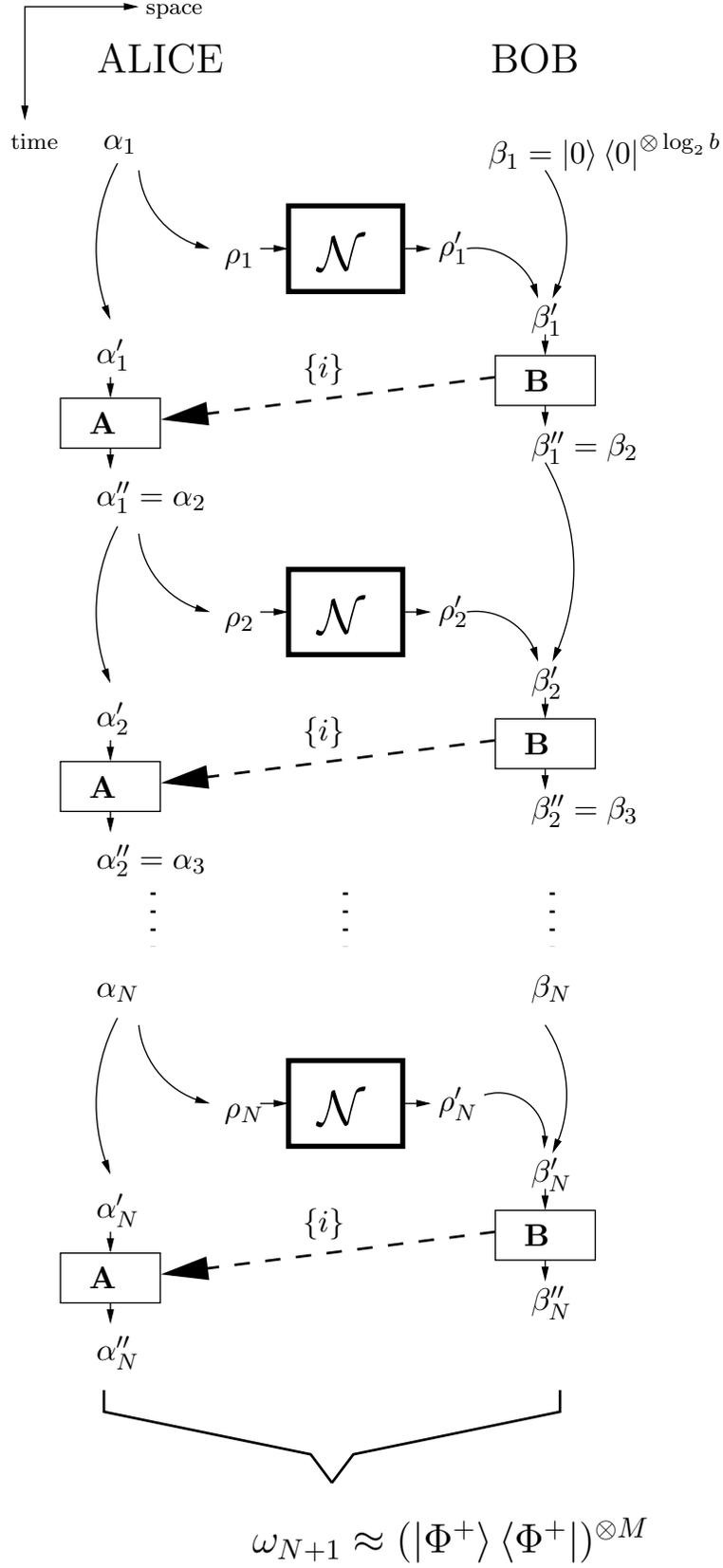}
\caption{\label{fig: Q_B-protocol}An $E_B$ protocols for channel
$\mathcal{N}$(See the text for details).}
\end{center}
\end{figure}

\noindent The second round of the protocols starts with Alice
holding $\alpha_2=tr_{(d_2\times b)}(\omega_2)$ and Bob holding
$\beta_2=tr_{(d_1^{N-1}\times a)}(\omega_2)$. After $N$ rounds of
protocols as seen in figure \ref{fig: Q_B-protocol}, we require
the fidelity between the quantum state shared between Alice and
Bob, $\omega_{N+1}$, and the quantum state,
$(\ket{\Phi^{+}}\bra{\Phi^{+}})^{\otimes M}$, to approach 1 as $N$
goes to infinity. Then we define $E_B(\mathcal{N})$ to be the
supremum of any attainable $\frac{M}{N\log_2(d_2)}$ - or simply
$M/N$ if $d_2=2$.

Note that in this work, when we discuss an $E_B$ protocol, for
brevity, we often say to compute the $E_B$ associated with the
protocol rather than to compute the lower bounds on
$E_B(\mathcal{N})$ impled by the protocol.

\subsection{$E_B\leq Q_2$}

To show $E_B\leq Q_2$, we simply convert any $E_B$ protocol to a
$Q_2$ protocol with the same rate. Suppose we have a protocol on
$\mathcal{N}$ and this $E_B(\mathcal{N})$ protocol achieves
$\frac{M}{N\log_2(d_2)}$, then at the end of this protocol Alice
and Bob share the quantum state
$\ket{\Phi^{+}}\bra{\Phi^{+}}^{\otimes M}$. Alice now uses the
forward classical communication channel to teleport any quantum
state $\rho\in\mathcal{H}_{2}^{\otimes M}$ and therefore this new
$Q_2(\mathcal{N})$ protocol achieves $\frac{M}{N\log_2(d_2)}$.

\subsection{$Q_B\leq E_B$}

This follows from the fact that $Q_B$ protocols are more
restricted than $E_B$ protocols because in defining quantum
capacities\cite{12} the sender is required to not only transmit
the quantum state $\rho$ but also preserve its entanglement with
the environment to which neither the sender nor the receiver has
access. In $E_B$ protocols, the sender is required to transmit
half of the maximally entangled states $\ket{\Phi^{+}}^M$ and is
in possession of the other half which she can manipulate in her
laboratory. Concisely, one can convert any $Q_B$ protocol to an
$E_B$ protocol as follows: Alice prepares
$\ket{\Phi^{+}}\bra{\Phi^{+}}^{\otimes M}\in
\mathcal{B}(\mathcal{H}_2^{\otimes M}\otimes
\mathcal{H}_2^{\otimes M})$ in her laboratory and performs the
$Q_B$ protocol on $\rho=(I/2)^{\otimes M}= tr_{(2^M)}
(\ket{\Phi^{+}}\bra{\Phi^{+}}^{\otimes M})$. At the end of the
protocol, Alice and Bob share the bipartite quantum state
$\ket{\Phi^{+}} \bra{\Phi^{+}}^{\otimes M}$ and hence
$E_B(\mathcal{N})\geq Q_B(\mathcal{N}) \geq
\frac{M}{N\log_2(d_2)}$.

\subsection{$E_B$ as quantum backward capacity with classical feedback}

In section \ref{sec: E_B-definition}, $E_B$ was defined as the
maximal asymptotic rate at which Alice shares the singlet state
$\ket{\Phi^{+}}$ with Bob with the assistance of a classical
feedback channel. Alternatively, we can associate $E_B$ with a
different operational meaning, namely the asymptotic rate at which
Bob can send quantum states to Alice. This is because after any
$E_B$ protocols Alice and Bob share the quantum states
$(\ket{\Phi^{+}}\bra{\Phi^{+}})^M$ and there is a classical
channel from Bob to Alice. Therefore, Bob can teleport any quantum
states $\rho\in\mathcal{H}_2^{\otimes M}$ to Alice and this
achieves the same yield $\frac{M}{N\log_2(d_2)}$ if we normalize
by the dimension of the output Hilbert space or if we assume the
input Hilbert space and the output Hilbert space are of the same
size. Trivially, if Bob can send quantum states to Alice, Bob can
choose to send half of the EPR pair $\ket{\Phi^{+}}$. Therefore
these two notions are equivalent to one another.

\section{Adaptive quantum error-correcting codes
(AQECC)}\label{sec: AQECC}

In quantum error-correcting codes\cite{N21,B5,N19,N20}, quantum
states are encoded into the subspace of some larger Hilbert space.
Although it has been discovered that quantum states can more
generally be encoded into a subsystem rather than a
subspace\cite{N23,N22}, we focus only on subspace encoding. Our
aim is to convert any quantum error-correcting codes (QECC) to new
adaptive $E_B$ protocols on the quantum depolarizing channel
$\mathcal{E}_p$. In section \ref{sec: stabilizer}, we briefly
review the stabilizer formalism; and in section \ref{sec:
E_B-protocol} we introduce the idea of AQECC. In the rest of the
section, we will illustrate with and compute the
$E_B(\mathcal{E}_p)$ for two QECC, namely the Cat code and Shor
code. We then consider how the recurrence methods - a 2-EPP - in
\ref{sec: recurrence-method-1} can be turned into an $E_B$
protocol. Finally we explain that the Leung-Shor method\cite{35}
in \ref{sec: LS} is in fact an $E_B$ protocol.

\subsection{Stabilizer formalism for QECC} \label{sec: stabilizer}

We will briefly review stabilizer formalism and introduce some
notations. A clear and detailed discussion can be found in
\cite{B5}. $G_n$ denotes the Pauli group on $n$ qubits, and
therefore consists of the n-fold tensor products of Pauli
matrices. For example,

$$G_1=\{\pm I, \pm i I, \pm X, \pm i X, \pm Y, \pm i Y, \pm Z, \pm i Z \}$$

\noindent where $X=\sigma_x$, $Y=\sigma_y$ and $Z=\sigma_z$. We
use subscripts to denote the qubit that a Pauli matrix acts on.
For example, $X_2Y_4$ means $I\otimes X \otimes I \otimes Y
\otimes I \otimes \ldots \otimes I \in G_n$. Generators of a
subgroup $S\subset G_n$ are independent if for any $i=1,2,3,\dots,
n-k$,

$$<g_1,\dots, g_{i-1},g_{i+1},\ldots,g_{n-k}>
\neq<g_1,\dots,g_{n-k}>.$$

\noindent We say a vector space $V_S\subset \mathcal{H}_2^{\otimes
n}$ is stabilized by a subgroup $S\subset G_n$ if for any
$\ket{\phi}\in V_S$ and for any $s\in S$,

$$s\ket{\phi}=\ket{\phi}.$$

\noindent The following lemma can be shown easily:

\begin{lem}
Let $S=<g_1,\ldots,g_{n-k}>$ be generated by $n-k$ independent and
commuting elements from $G_n$, and $-I\not\in S$. Then $V_S$ is a
$2^k$-dimensional vector space.
\end{lem}

Therefore to specify a $2^k$-dimensional subspace for
error-correcting codes, we only need to specify $n-k$ independent
generators $g_1,\ldots,g_{n-k}$. However we still need to specify
the logical basis vectors $\ket{x_1,\ldots,x_k}_L$ within $V_S$.
In this work, we only deal with codes where $k=1$. Therefore, it
suffices to specify the logical $\bar{X}$ and logical $\bar{Z}$
such that $\bar{X}\ket{0}_L=\ket{1}_L \in \mathcal{H}_2^{\otimes
n}$, $\bar{X}\ket{1}_L=\ket{0}_L \in \mathcal{H}_2^{\otimes n}$,
$\bar{Z}\ket{0}_L=\ket{0}_L \in \mathcal{H}_2^{\otimes n}$ and
$\bar{Z}\ket{1}_L=- \ket{1}_L \in \mathcal{H}_2^{\otimes n}$. Note
that in doing so, we indirectly specify $\ket{0}_L$ and
$\ket{1}_L$.

\subsection{$E_B$ protocols via AQECC}\label{sec: E_B-protocol}

Recall the aim of any $E_B$ protocols is for Alice to share the
bipartite state $\ket{\Phi^{+}}=\frac{1}{\sqrt{2}}(\ket{00}
+\ket{11})$ with Bob. We will explain our idea of turning a QECC
to an $E_B$ protocol in two steps.

The first step is to simply encode half of the EPR pair
$\ket{\Phi^{+}}$ in an $[n,1]$ stabilizer code, one that encodes a
qubit in an $2^n$-dimensional Hilbert space
$\mathcal{H}_2^{\otimes n}$. Alice performs the encoding

{\setlength\arraycolsep{2pt}
\begin{eqnarray}
\mathbf{A}: \mathcal{B}(\mathcal{H}_2) \longrightarrow
\mathcal{B}(\mathcal{H}_2^{\otimes n})
\nonumber\\
tr_2(\ket{\Phi^{+}}\bra{\Phi^{+}}) \mapsto  \alpha_1 \nonumber
\end{eqnarray}}

\noindent and then sends the $n$ qubits through the
$p$-depolarizing channel

{\setlength\arraycolsep{2pt}
\begin{eqnarray}
\mathcal{E}_p: \mathcal{B}(\mathcal{H}_2) &\longrightarrow &
\mathcal{B}(\mathcal{H}_2) \nonumber\\
\rho &\mapsto & \frac{1+3p}{4} \times \rho +
\frac{1-p}{4}\times(\sigma_x \rho \sigma_x^{\dagger}+\sigma_y \rho
\sigma_y^{\dagger}+\sigma_z \rho \sigma_z^{\dagger}).\nonumber
\end{eqnarray}}

\begin{figure}
\begin{center}
\includegraphics[width=150mm]{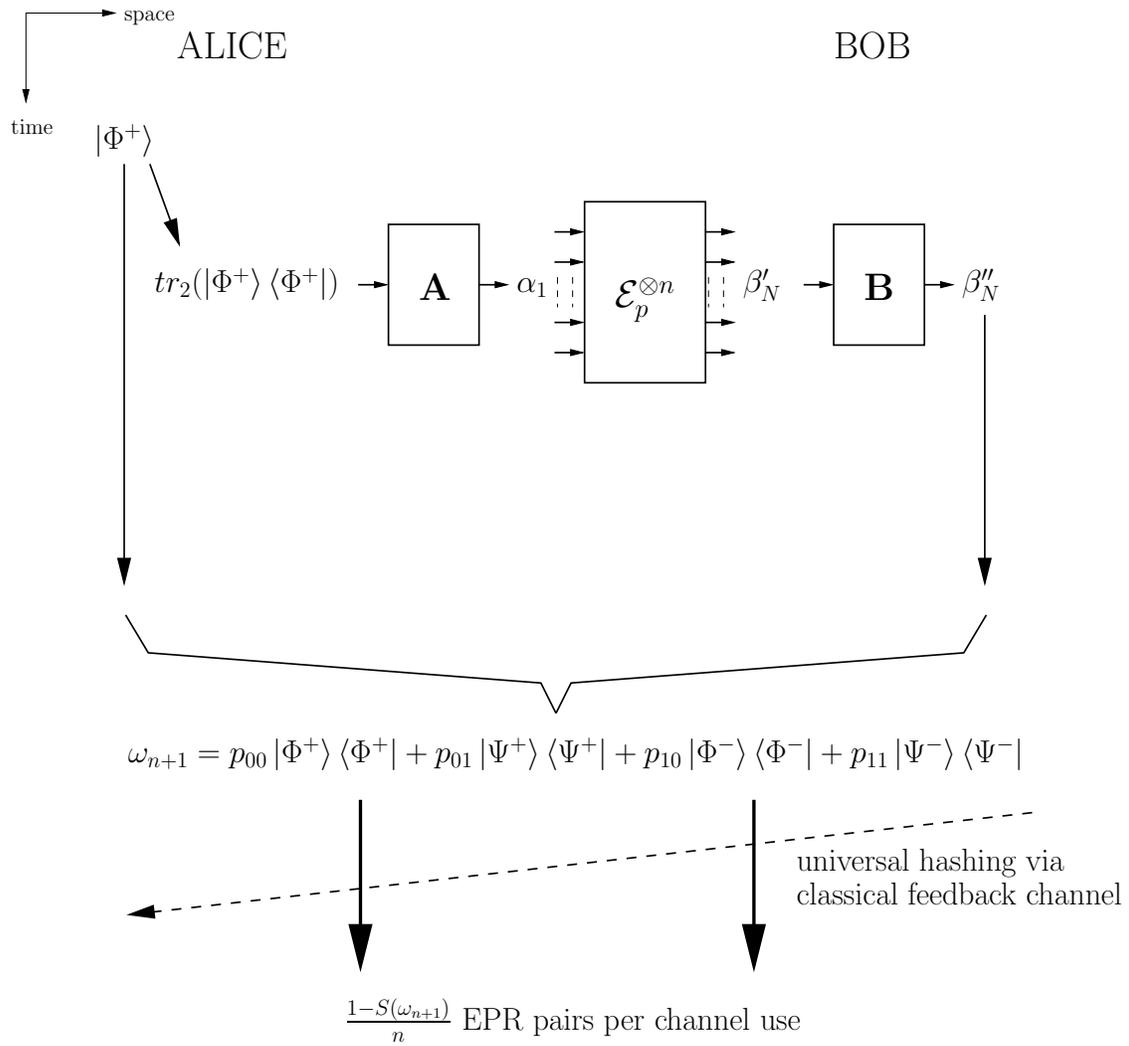}
\caption{\label{fig: QECC-protocol-1}Encoding half of the EPR pair
$\ket{\Phi^{+}}$ with a QECC .}
\end{center}
\end{figure}

\noindent Since the error elements of the $p$-depolarizing channel
are Pauli matrices, Alice can choose the logical basis states (or
alternatively the logical operators $\bar{X}, \bar{Z}$ as we
explained in the previous section) in such a way that after the
error-correction operation $\mathbf{B}$, the encoded qubit has
either an $X$ error, a $Y$ error, a $Z$ error or no error. Since
$X\ket{\Phi^{+}}=\ket{\Psi^{+}}$, $Y\ket{\Phi^{+}}=\ket{\Psi^{-}}$
and $Z\ket{\Phi^{+}}=\ket{\Phi^{-}}$, the bipartite state between
Alice and Bob will be a probabilistic mixture of the four Bell
states. Therefore Bob can use the classical feedback channel to
perform universal hashing and distill perfect EPR pairs
$\ket{\Phi^{+}}=\frac{1}{\sqrt{2}}(\ket{00} +\ket{11})$. This
first step is illustrated in figure \ref{fig: QECC-protocol-1}.

The second step is to modify what has just been described so as to
achieve a higher rate. Recall an $[n,1]$ stabilizer code is
described by the generators of a subgroup $S=<g_1,g_2,\ldots,
g_{n-2}, g_{n-1}>$. The error-correcting operation $\mathbf{B}$
performed by Bob involves measuring the observables $g_1, g_2,
\ldots, g_{n-1}$ since they are all tenser products of Pauli
matrices acting on n qubits. Note that, however, many of the
$g_i$'s have identity action on all but a few qubits. For example,
in 9-bit Shor code, $g_1=Z_1Z_2$($=Z_1\otimes Z_2 \otimes I_3
\otimes I_4 \otimes I_5 \otimes I_6 \otimes I_7 \otimes I_8
\otimes I_9$). Also, whenever a measurement result `-1' is
obtained, it means some errors have occurred. In the case of Shor
code, if Bob takes a measurement on the first two qubits
immediately after he receives them from Alice and the measurement
result is `-1', it is better for Bob to use the classical feedback
channel to inform Alice that some errors have occurred in the
first 2 qubits and they should give up this block of transmission
and start all over. It is because the quantum state $\omega_{n+1}$
Alice and Bob obtained after n channel uses and decoding will be
more mixed - or in other words of higher entropy - if some errors
have occurred. It is thus more economical to not continue with
this particular block of codes and give up the few qubits that
have already been transmitted.

It is thus important to arrange the order of the measurements
$g_1, g_2, \ldots, g_{n-1}$ such that it only involves as few more
qubits as possible when one goes down the list. So that when an
error is detected early on, Alice and Bob can stop the block and
start all over so as to save more channel uses. For example, the
generators of Shor code can be arranged as follows:

{\setlength\arraycolsep{2pt}
\begin{eqnarray}
g_1 & = & Z_1\otimes Z_2 \otimes I_3 \otimes I_4 \otimes I_5
\otimes I_6 \otimes I_7 \otimes I_8 \otimes I_9 \nonumber\\
g_2 & = & I_1\otimes Z_2 \otimes Z_3 \otimes I_4 \otimes I_5
\otimes I_6 \otimes I_7 \otimes I_8 \otimes I_9 \nonumber\\
g_3 & = & I_1\otimes I_2 \otimes I_3 \otimes Z_4 \otimes Z_5
\otimes I_6 \otimes I_7 \otimes I_8 \otimes I_9 \nonumber\\
g_4 & = & I_1\otimes I_2 \otimes I_3 \otimes I_4 \otimes Z_5
\otimes Z_6 \otimes I_7 \otimes I_8 \otimes I_9 \nonumber\\
g_5 & = & X_1\otimes X_2 \otimes X_3 \otimes X_4 \otimes X_5
\otimes X_6 \otimes I_7 \otimes I_8 \otimes I_9 \nonumber\\
g_6 & = & I_1\otimes I_2 \otimes I_3 \otimes I_4 \otimes I_5
\otimes I_6 \otimes Z_7 \otimes Z_8 \otimes I_9 \nonumber\\
g_7 & = & I_1\otimes I_2 \otimes I_3 \otimes I_4 \otimes I_5
\otimes I_6 \otimes I_7 \otimes Z_8 \otimes Z_9 \nonumber\\
g_8 & = & I_1\otimes I_2 \otimes I_3 \otimes X_4 \otimes X_5
\otimes X_6 \otimes X_7 \otimes X_8 \otimes X_9 \label{eqn:
shor-code-generators}
\end{eqnarray}}

\noindent It is conceivable that after a large portion of the
qubits in a block have been transmitted, it is better to continue
even if an error is detected. It is indeed the case for Shor code
when the probability parameter $p$ of the channel $\mathcal{E}_p$
is large. In the next two sections, we will apply this AQECC idea
to Cat code and Shor code, and compute the lower bounds on
$E_B(\mathcal{E}_p)$ implied by these codes.

\subsection{Cat code and modified Cat code} \label{sec: cat-code}

The n-bit Cat (repetition) code is an $[n,1]$ stabilizer code with
the following generators

{\setlength\arraycolsep{2pt}
\begin{eqnarray}
g_1 & = & Z_1Z_2 \nonumber\\
g_2 & = & Z_2Z_3 \nonumber\\
g_3 & = & Z_3Z_4 \nonumber\\
\vdots & \vdots & \vdots \nonumber\\
g_{n-2} & = & Z_{n-2}Z_{n-1}\nonumber\\
g_{n-1} & = & Z_{n-1}Z_n \nonumber
\end{eqnarray}}

\noindent and we choose the following logical operators

{\setlength\arraycolsep{2pt}
\begin{eqnarray}
\bar{X} & = & X_1X_2\ldots X_{n-1}X_{n} \nonumber\\
\bar{Z} & = & Z_1Z_2\ldots Z_{n-1}Z_{n} \textrm{ if n is odd and
} \nonumber \\
\bar{Z} & = & Z_1Z_2\ldots Z_{n-1}I_{n} \textrm{ if n is even.}
\nonumber
\end{eqnarray}}

\noindent This in turn determines the logical computational basis

$$\ket{0}_L=\ket{00\ldots 00} \in \mathcal{H}_2^{\otimes n} \textrm{ and } \ket{1}_L
=\ket{11\ldots 11} \in \mathcal{H}_2^{\otimes n}.$$

\noindent Therefore, the singlet state $\frac{1}{\sqrt{2}}
(\ket{00}+\ket{11})\in\mathcal{H}_2^{\otimes 2}$ is encoded as
$\frac{1}{\sqrt{2}}(\ket{00\ldots 00}+\ket{11\ldots
11})\in\mathcal{H}_2^{\otimes n+1}$ in Alice's laboratory. Alice
will send the last n qubits to Bob via the channel
$\mathcal{E}_p$. In accordance with the AQECC idea in the previous
section, Alice sends the first two qubits first and Bob takes the
measurement $g_1$. If the measurement result is `-1', Bob will
inform Alice of the result via the classical feedback channel and
Alice will discard the n-1 qubits remaining in her laboratory and
start all over by encoding another EPR pair and sending the
quantum states. If the measurement result is `+1', Bob will inform
Alice of the result and Alice will continue to send the third
qubit. Bob will then measure $g_2$. This continues until all n
qubits are passed to Bob and Bob gets `+1' in all n-1 measurements
$g_1,g_2,\ldots,g_{n-1}$. Alice and Bob will then process a
bipartite quantum state $\omega_{n+1}\equiv
p_{00}\ket{\Phi^+}\bra{\Phi^+} + p_{01}\ket{\Psi^+}\bra{\Psi^+} +
p_{10}\ket{\Phi^-}\bra{\Phi^-} + p_{11}\ket{\Psi^-}\bra{\Psi^-}$
that is Bell diagonal. If Alice and Bob repeat the process until
they share N copies of $\omega_{n+1}$, i.e. $\omega_{n+1}^{\otimes
N}$, they can perform universal hashing on these states and they
will have $N\Big(1-H(p_{00},p_{01},p_{10},p_{11})\Big)$ EPR pairs
$\ket{\Phi^+}$. However we are interested in the yield per channel
use. Let $p_i=\textrm{prob}(\textrm{`+1' for measurement } g_i)$.
Then the average number of channel uses needed before we
successfully pass a block of n-qubit Cat code through the
depolarizing channel is given by

{\setlength\arraycolsep{2pt}
\begin{eqnarray}
n^*&= & \bigg( \sum_{i=2}^{n-1} \Big(i \times (\prod_{j=1}^{i-2}
p_j) \times (1-p_{i-1}) \Big) + n\times \prod_{i=1}^{n-2} p_i
\bigg) / \bigg(n\times \prod_{i=1}^{n-1}p_i \bigg)  \nonumber \\
& = & \bigg( 2\times (1-p_1) + 3\times p_1 \times (1-p_2)+ \ldots
+ (n-1)\times p_1\times p_2\times \ldots p_{n-3} \times (1-
p_{n-2}) \nonumber\\
&& + n\times p_1\times p_2 \times \ldots \times p_{n-2} \bigg)/
\bigg( n \times p_1\times \ldots \times p_{n-1}\bigg). \nonumber
\end{eqnarray}}

\noindent From this, the number of EPR pairs per channel use is

{\setlength\arraycolsep{2pt}
\begin{eqnarray}
&& \frac{1}{N \times n^* \times n}\times
N\Big(1-H(p_{00},p_{01},p_{10},p_{11})\Big)\nonumber \\
&= & \frac{\bigg(\prod_{i=1}^{n-1}p_i\bigg)\times
\bigg(1-H(p_{00},p_{01},p_{10},p_{11})\bigg)}
{\bigg(\sum_{i=2}^{n-1} \Big(i \times (\prod_{j=1}^{i-2} p_j)
\times (1-p_{i-1}) \Big) + n\times \prod_{i=1}^{n-2} p_i \bigg)
}.\label{eqn: Cat-yield}
\end{eqnarray}}

\begin{figure}
\begin{center}
\includegraphics[height=60mm]{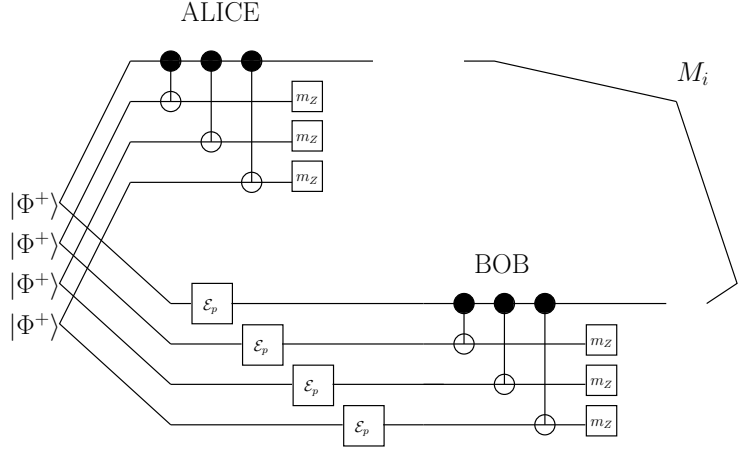}
\caption{\label{fig: Cat-protocol}4-bit Cat code in the language
of entanglement purification protocols. Note that in our protocols
if Bob's measurement results do not agree with Alice's, then not
all qubits will be sent through $\mathcal{E}_p$. Alice's
measurement results are assumed to be all `+1' so Alice need not
send Bob any classical information even though Bob `compares' his
results against Alice's(See the text for details).}
\end{center}
\end{figure}

\noindent We now present how to calculate the probabilities $p_1,
\ldots, p_{n-1}$ and the quantum state
$\omega_{n+1}=p_{00}\ket{\Phi^+}\bra{\Phi^+} +
p_{01}\ket{\Psi^+}\bra{\Psi^+} + p_{10}\ket{\Phi^-}\bra{\Phi^-} +
p_{11}\ket{\Psi^-}\bra{\Psi^-}$. The computation can be given by a
simple recurrence relation \cite{50,49} which can be understood
more easily in the language of entanglement purification
protocols. Owing to the formal equivalence between measuring half
of a Bell state and preparing a qubit, the encoding and decoding
of the Cat code can be viewed as a 1-EPP as shown in figure
\ref{fig: Cat-protocol} for $n = 4$. Note that in order for the
purification protocols to work, it appears Alice has to send her
measurement results to Bob via a side forward communication
channel as in 2-EPP. This is in fact not the case because even
though the measurement results are non-deterministic, Alice can
perform the measurements before she sends the 4 qubits (or
generally n qubits). One can pretend Alice takes measurements for
as many times as needed until she gets all `+1' before she sends
the other halves of the quantum states via $\mathcal{E}_p$.
Therefore Alice need not tell Bob the results because Bob already
knew the results were all `+1'. (Of course, in reality, Alice can
apply unitary operation in her laboratory to transform the states
to what she needs even if some measurement results are `-1'.)

\begin{figure}
\begin{center}
\centerline{\includegraphics[width=120mm]{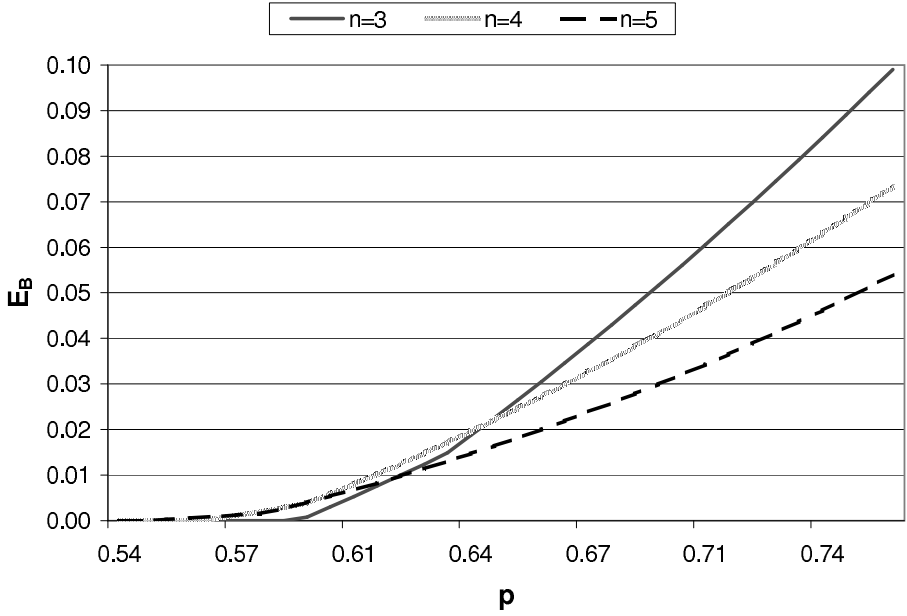}}
\caption{\label{fig: Cat-yield}Lower bounds on
$E_B(\mathcal{E}_p)$ via n-bit Cat code and modified Cat code. See
the text for details.}
\end{center}
\end{figure}

Note that applying a CNOT gate on the first and the (i-1)th qubits
followed by measuring the (i-1)th qubit along the z-axis as shown
in figure \ref{fig: Cat-protocol} is the same as measuring $g_i$,
and we are interested in keeping track of the quantum state of the
first qubit that passed through $\mathcal{E}_p$ after each
measurement $g_i$. We are only interested in its quantum state if
the measuring result is `+1', since we otherwise discard the
states and start all over. Denote this state by $M_i$, and we have
the following relations \cite{49}:

$$p_{i+1} =  (F+G)\bra{\Phi^+} M_i \ket{\Phi^+} +
(2G)\bra{\Psi^+} M_i \ket{\Psi^+} +(F+G)\bra{\Phi^-} M_i
\ket{\Phi^-} + (2G)\bra{\Psi^-} M_i \ket{\Psi^-} $$

\begin{figure}
\begin{center}
\centerline{\includegraphics[width=120mm]{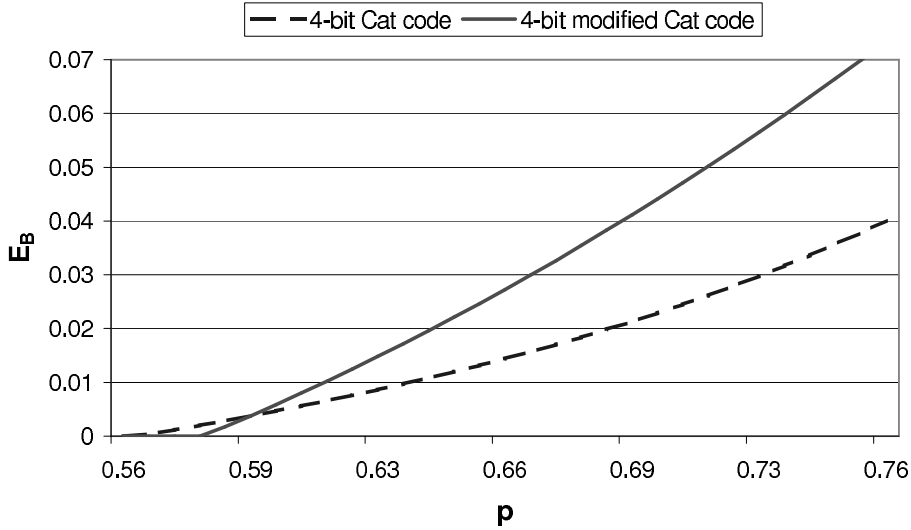}}
\caption{\label{fig: mod-cat-vs-cat-yield}4-bit Cat code vs. 4-bit
modified Cat code.}
\end{center}
\end{figure}

{\setlength\arraycolsep{1pt}
\begin{eqnarray}
\bra{\Phi^+} M_{i+1} \ket{\Phi^+} & = & \frac{F\bra{\Phi^+} M_i
\ket{\Phi^+} + G\bra{\Phi^-} M_i \ket{\Phi^-}} {p_i} \nonumber \\
\bra{\Psi^+} M_{i+1} \ket{\Psi^+} & = & \frac{G\bra{\Psi^+} M_i
\ket{\Psi^+} + G\bra{\Psi^-} M_i \ket{\Psi^-}}{p_i} \nonumber \\
\bra{\Phi^-} M_{i+1} \ket{\Phi^-} & = & \frac{G\bra{\Phi^+} M_i
\ket{\Phi^+} + F\bra{\Phi^-} M_i \ket{\Phi^-}}{p_i} \nonumber \\
\bra{\Psi^-} M_{i+1} \ket{\Psi^-} & = & \frac{G\bra{\Psi^+} M_i
\ket{\Psi^+}+ G\bra{\Psi^-} M_i \ket{\Psi^-}}{p_i}  \nonumber
\end{eqnarray}}

\noindent where $F=\frac{3p+1}{4}$, $G=\frac{1-F}{3}$ and
$M_{n-1}=\omega_{n+1}$. From these equations and (\ref{eqn:
Cat-yield}), we compute the lower bounds on $E_B$ with n-bit Cat
code and modified Cat code for $n = 3,4,5$ in figure \ref{fig:
Cat-yield}. Modified Cat code differs from Cat code in the same
way that the modified recurrence method differs from the
recurrence method. Namely, Bob switches the
$\ket{\Phi^-}\bra{\Phi^-}$ and $\ket{\Psi^-}\bra{\Psi^-}$
components in the probabilistic mixture of Bell states after each
measurement. This can be done by first applying a bilateral
$\pi/2$ rotation $B_x$ and then a unilateral $\pi$ rotation
$\sigma_x$ \cite{35}. Modified Cat code outperforms Cat code when
the channel is less noisy(large p), but Cat code performs slightly
better when the channel is very noisy and hence achieves a lower
threshold value. In figure \ref{fig: mod-cat-vs-cat-yield}, we
plot the yield for 4-bit Cat code and modified Cat code
separately.

\subsection{Shor code}\label{sec: shor-code}

The generators of Shor code are listed in (\ref{eqn:
shor-code-generators}). The logical operators and logical
computational basis states are as follows:

{\setlength\arraycolsep{2pt}
\begin{eqnarray}
\bar{X}&=&Z_1Z_2Z_3Z_4Z_5Z_6Z_7Z_8Z_9\nonumber\\
\bar{Z}&=&X_1X_2X_3X_4X_5X_6X_7X_8X_9\nonumber
\end{eqnarray}}

{\setlength\arraycolsep{2pt}
\begin{eqnarray}
\ket{0}_L&=&\frac{(\ket{000}+\ket{111})(\ket{000}+\ket{111})(\ket{000}+\ket{111})}{2\sqrt{2}}\nonumber\\
\ket{1}_L&=&\frac{(\ket{000}-\ket{111})(\ket{000}-\ket{111})(\ket{000}-\ket{111})}{2\sqrt{2}}.\nonumber
\end{eqnarray}}

\noindent As aforementioned, for 9-bit Shor code, the optimal
AQECC protocols are slightly different for different levels of
noise. We can divide the protocols into 3 regions:
\\
\\\begin{tabular}{|c|c|}
  \hline
  p & protocol \\
  \hline
  less than 0.75 & start all over if any measurement result is `-1' \\
  \hline
  between 0.75 & start all over if any of the first 7 measurement results is `-1'; \\
  and 0.78 &otherwise continue with the regular error-correcting operation\\
  \hline
  great than 0.78 & start all over if any of the first 4 measurement results is `-1';\\
    &otherwise continue with the regular error-correcting operation\\
  \hline
\end{tabular}

\noindent \\In the first region (p less than 0.75), one only has
to enumerate all $4^9$ error possibilities in the 9 channel uses
and adds up all probabilities associated with having an $X$ error,
a $Y$ error, a $Z$ error or no error on the encoded qubit. Then
the $E_B$ rate achieved for $\mathcal{E}_p$ is given by:

$$\frac{p_1\times p_2\times \ldots \times p_8 \times
\big(1-H(p{00},p{01},p{10},p{11})\big)}{n^*}$$

\noindent where

{\setlength\arraycolsep{2pt}
\begin{eqnarray}
n^*&=&2\times (1-p_1) + 3\times p_1(1-p_2) + 5\times p_1p_2(1-p_3)
+ 6\times p_1p_2p_3(1-p_4)\nonumber\\
&&+ 6 \times p_1p_2p_3p_4(1-p_5)+8\times
p_1p_2p_3p_4p_5(1-p_6)+9\times p_1p_2p_3p_4p_5p_6.\nonumber
\end{eqnarray}}

\begin{figure}
\begin{center}
\centerline{\includegraphics[width=140mm]{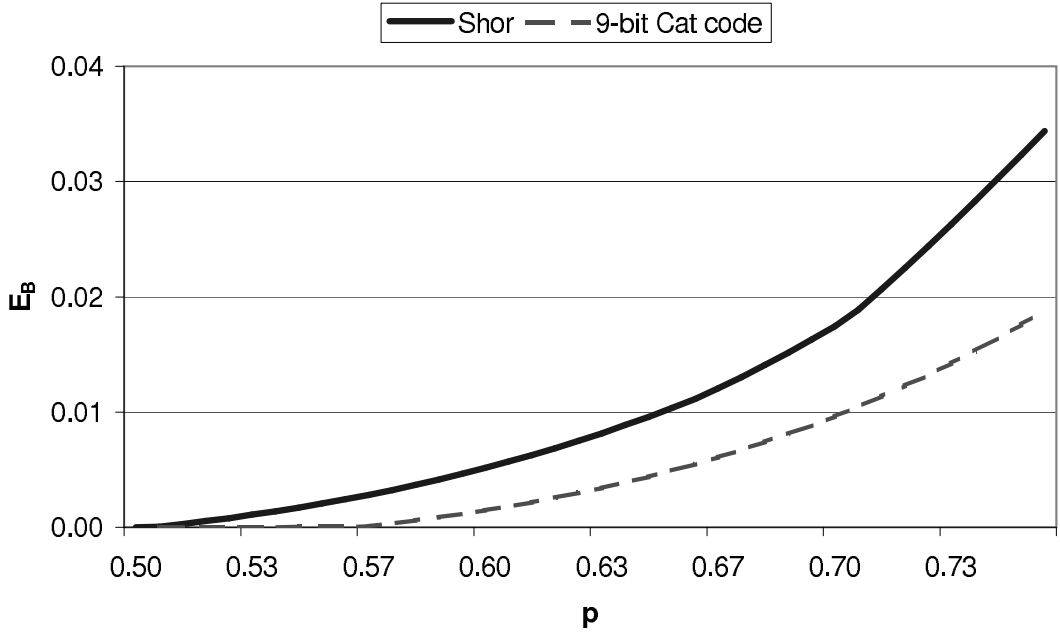}}
\caption{\label{fig: shor-yield}Lower bounds on
$E_B(\mathcal{E}_p)$ via 9-bit Shor code and 9-bit Cat code.}
\end{center}
\end{figure}

\noindent In the second and the third region, the computation is
slightly different. We will illustrate with the third region, and
the computation for the second region is similar. Since Alice and
Bob will start all over if any of the first 4 measurement results
is `-1', there are only $2^{(8-4)}=16$ possible measurement
results given that the whole block of 9 qubits were sent through
the channel. Denote the 4-tuple measurement results by
$m\in\{0,1,\ldots,15\}$. For each measurement result, Bob will
carry out error-correcting operation as in the standard 9-bit Shor
code and inform Alice which of the 16 measurement results this
block of 9 qubits has. Then after a large number of 9-bit blocks
are transmitted successfully, Alice and Bob share a large number
of each of the 16 types of Bell-diagonal probabilistic mixtures so
that they can perform universal hashing on each of these 16 types
of mixtures separately. And the $E_B$ rate achieved is given by

$$\sum_{m} \Bigg( \frac{1}{n^{**}}\textrm{ prob(measurement result is m) }
\bigg(1-H(p{00},p{01},p{10},p{11}|m)\bigg)\Bigg)$$

\noindent where $H(p{00},p{01},p{10},p{11}|m)$ is the entropy of
the probabilistic mixture given a particular measure result
$m\in\{0,1,\ldots,15\}$ has occurred and $n^{**}=2\times(1-p_1) +
3\times p_1(1-p_2) + 5\times p_1p_2(1-p_3)+ 6\times
p_1p_2p_3(1-p_4) +9\times p_1p_2p_3p_4$. In figure \ref{fig:
shor-yield}, we plot the $E_B$ rate achieved; for comparison $E_B$
rate achieved for 9-bit Cat code is also shown.

\subsection{Modified recurrence method}

\begin{figure}
\begin{center}
\centerline{\includegraphics[width=110mm]{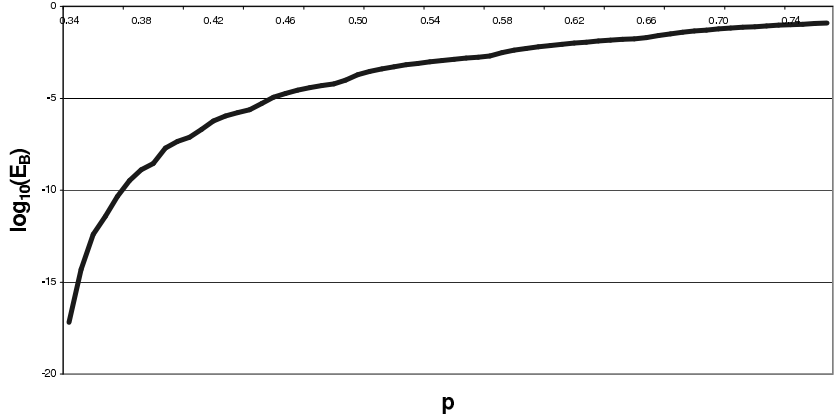}}
\caption{\label{fig: recurrence-yield}Lower bounds on
$E_B(\mathcal{E}_p)$ via the modified recurrence method.}
\end{center}
\end{figure}

Modified recurrence method as described in \cite{35} is a 2-EPP
which requires two-way classical communication. Although Alice can
perform the measurement before she sends halves of the EPR pairs
$\ket{\Phi^+}$ through $\mathcal{E}_p$ so that Bob need not know
her measurement results in the first round, as we discussed in
section \ref{sec: cat-code} and \ref{sec: shor-code}, an iterative
process is not possible. In particular, one round of recurrence
plus universal hashing via the classical feedback channel achieve
positive $E_B$ rate only for $p>0.638$. If Alice and Bob want to
carry out another round of the modified recurrence method, she
needs a forward channel to communicate her measurement results to
Bob. Since the only forward channel for Alice is $\mathcal{E}_p$,
a straightforward extension, therefore, is to use the channel
$\mathcal{E}_p$ to send her measurement results to Bob. As a
result, from the second round onwards, one classical bit per pair
is required for each round of recurrence.

By proving the additivity conjecture for the quantum depolarizing
channel $\mathcal{E}_p$, the formula for the classical capacity of
$\mathcal{E}_p$ is known\cite{13}:

$$C(\mathcal{E}_p)= 1 + \Big( \frac{1+p}{2}\Big) \log_2
\Big( \frac{1+p}{2}\Big) + \Big(\frac{1-p}{2}\Big)\log_2
\Big(\frac{1-p}{2}\Big)= 1 - H\Big(\frac{1+p}{2},
\frac{1-p}{2}\Big).\label{eqn: depolarizing-capacity}$$

\noindent Then the $E_B$ yield implied by this method for $k$
rounds of recurrence before switching to universal hashing is
given by:

{\setlength\arraycolsep{2pt}
\begin{eqnarray}
\Bigg(\frac{p_{pass}^{(1)}}{2}\Bigg)\times
\Bigg(\frac{p_{pass}^{(2)}}{2+1/C(\mathcal{E}_p)}\Bigg)\times
\ldots\Bigg(\frac{p_{pass}^{(k)}}{2+1/C(\mathcal{E}_p)}\Bigg)
\times \Bigg( 1 - H\Big(p_{00}^{(k)}, p_{01}^{(k)}, p_{10}^{(k)},
p_{11}^{(k)}\Big)\Bigg)\nonumber
\end{eqnarray}}

\noindent where $p_{00}^{(k)}$, $p_{01}^{(k)}$, $p_{10}^{(k)}$,
$p_{11}^{(k)}$ and $p_{pass}^{(i)}$ for $i=1,2,\ldots, k$ are
given by the recurrence relations (\ref{eqn: recurrence-method1})
and (\ref{eqn: recurrence-method2}) in section \ref{sec:
recurrence-method-1}. In figure \ref{fig: recurrence-yield}, we
plot the $E_B$ rate achieved by this method.

\subsection{Leung-Shor method}

\begin{figure}
\begin{center}
\centerline{\includegraphics[width=90mm]{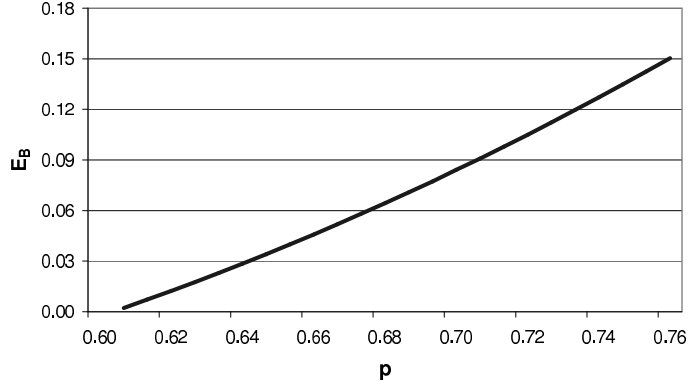}}
\caption{\label{fig: 3-LS-yield}Lower bounds on
$E_B(\mathcal{E}_p)$ via the Leung-Shor method.}
\end{center}
\end{figure}

The Leung-Shor method\cite{LS1} introduced in section \ref{sec:
LS} is in fact an $E_B$ protocol. Alice only needs to encode the
qubits into what they would have been if the measurement results
in figure \ref{fig: 2-LS} were both `+1'. In figure \ref{fig:
3-LS-yield}, we plot the $E_B$ rate achieved. In figure \ref{fig:
3-E_B-yield}, we compare the yield of the four methods in this
section.

\begin{figure}[h]
\begin{center}
\centerline{\includegraphics[width=130mm]{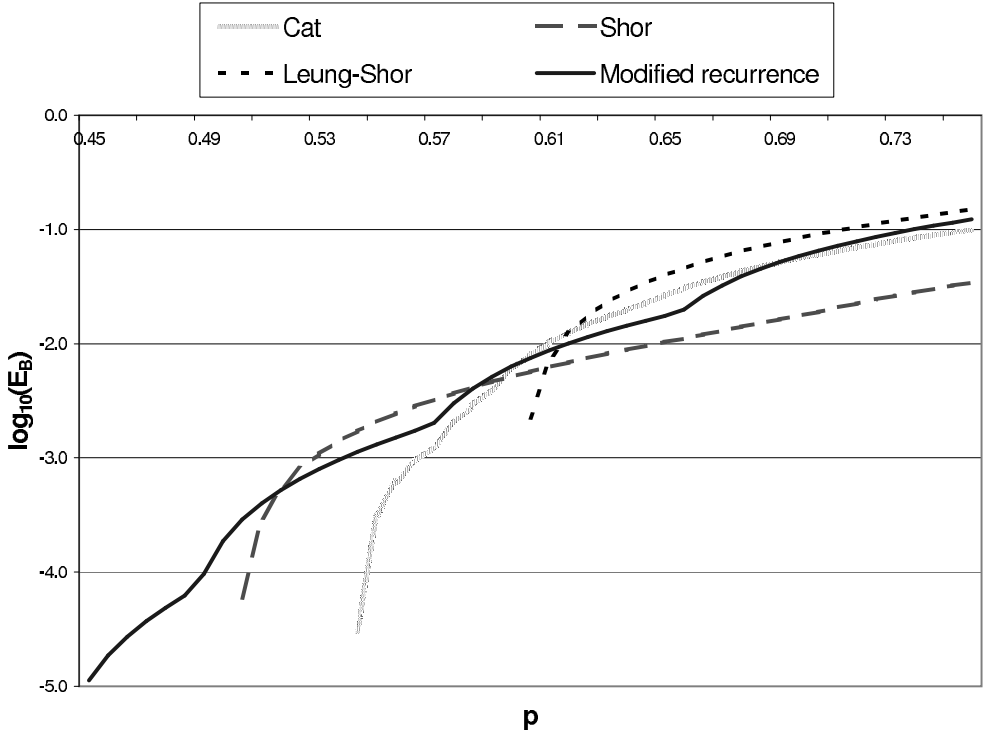}}
\caption{\label{fig: 3-E_B-yield}Lower bounds on
$E_B(\mathcal{E}_p)$.}
\end{center}
\end{figure}

\section{New lower bounds on $Q_B$} \label{sec: new-Q_B}

We will establish the following lemma which gives lower bounds on
$Q_B$ based on $E_B$ protocols:

\begin{lem}
{\setlength\arraycolsep{2pt}
\begin{eqnarray}
Q_B(\mathcal{E}_p) & \geq &
\frac{C(\mathcal{E}_p)}{1+\frac{C(\mathcal{E}_p)}{E_B(\mathcal{E}_p)}}\nonumber
\end{eqnarray}}
\noindent where $C(\mathcal{E}_p)= 1 - H\big(\frac{1+p}{2},
\frac{1-p}{2}\big).$
\end{lem}

\textbf{\emph{Proof}} In an $E_B$ protocol, Alice and Bob share M
EPR pairs $\ket{\Phi^{+}}$ in N channel uses. Therefore,
$E_B(\mathcal{E}_B)=M/N$. To teleport a quantum state
$\rho\in\mathcal{H}_2^{\otimes M}$, Alice can use the channel
$\mathcal{E}_p$ for $\frac{M}{C(\mathcal{E}_p)}$ many times to
send M bits of classical information to Bob. Thus,

{\setlength\arraycolsep{2pt}
\begin{eqnarray}
Q_B(\mathcal{E}_p) & \geq & \frac{M}{N+\frac{M}{C(\mathcal{E}_p)}}
=\frac{M/N}{1+\frac{M/N}{C(\mathcal{E}_p)}}\nonumber \\
&=&\frac{E_B(\mathcal{E}_p)}{1+\frac{E_B(\mathcal{E}_p)}{C(\mathcal{E}_p)}}
=\frac{C(\mathcal{E}_p)}{1+\frac{C(\mathcal{E}_p)}{E_B(\mathcal{E}_p)}}.\nonumber
\end{eqnarray}}
\hfill$\square$

\begin{figure}
\begin{center}
\centerline{\includegraphics[width=120mm]{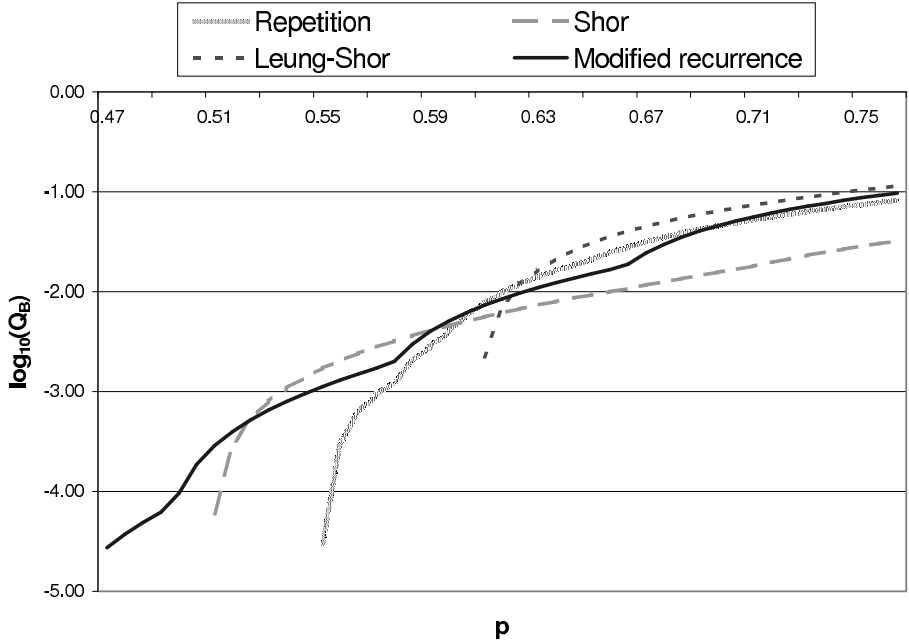}}
\caption{\label{fig: 3-Q_B-yield}Lower bounds on
$Q_B(\mathcal{E}_p)$.}
\end{center}
\end{figure}

\noindent From the lemma, any lower bounds on $E_B$ will imply
lower bounds on $Q_B$. The lower bounds are presented in figure
\ref{fig: 3-Q_B-yield}.

\section{Threshold of Cat code}\label{sec: threshold}

\begin{figure}
\begin{center}
\centerline{\includegraphics[width=130mm]{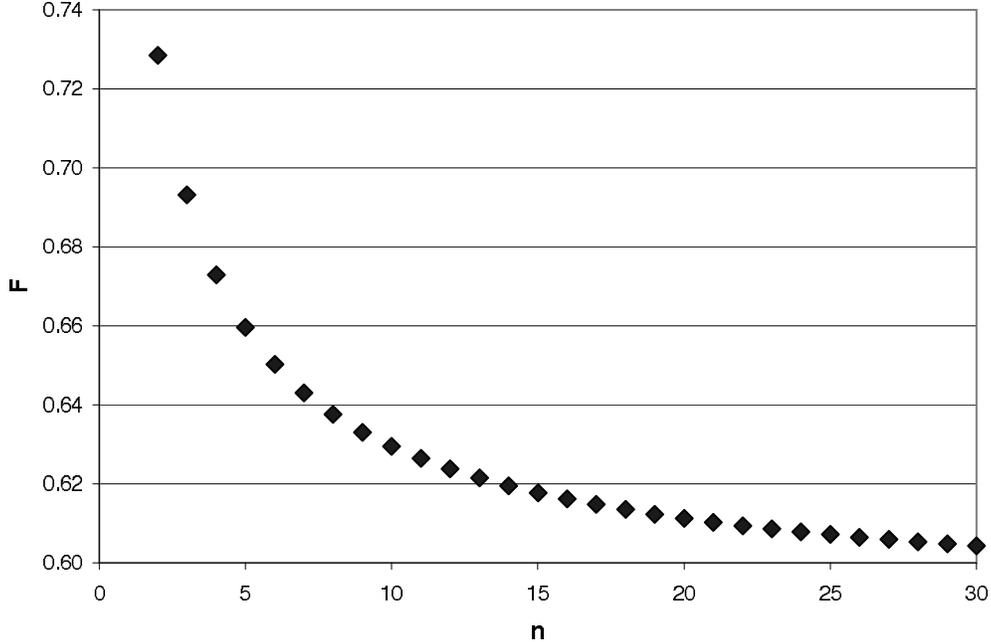}}
\caption{\label{fig: 3-threshold}Threshold fidelity
$F=\frac{3p+1}{4}$ for n-bit Cat code.}
\end{center}
\end{figure}

It has been shown that in the absence of side classical
communication one can achieve non-zero capacity for lower
threshold fidelity $F=\frac{3p+1}{4}$ by concatenating 5-bit Cat
code inside a random code (hashing)\cite{50}. Threshold fidelity
for concatenating n-bit Cat code into random code was also
studied. It was found that threshold fidelities fall into two
smooth curves, one for even n and one for odd n, but both curves
increase with n, i.e. one does not attain lower threshold by using
a longer Cat code. We therefore compute the threshold fidelity for
n-bit Cat code in figure \ref{fig: 3-threshold} and we found that
these phenomena do not occur in AQECC.

\section{Discussion on $Q_B$, $E_B$ and $Q_2$}

In this work, we defined the quantum entanglement capacity with
classical feedback $E_B$ for any quantum discrete memoryless
channel. For any channel, this quantity was shown to lie between
two other capacities, namely the quantum capacity with classical
feedback $Q_B$ and the quantum capacity with two-way classical
communication $Q_2$. It is an open question whether there exists
quantum channel for which $Q_2>Q_B$. While the introduction of
this new, intermediate quantity $E_B$ does not simplify the
question, it is our hope to shed some light on and provide other
means to tackle this open problem. In section \ref{sec:
E_B-between}, we provided an alternate operational interpretation
of this quantity: it represents the amount of quantum information
Bob can send to Alice. It is our hope that, by working with this
interpretation, one might be able to prove a non-trivial upper
bound on $E_B$ and hence lead to a separation between $Q_B$ and
$Q_2$.

We converted many of the well-known QECC into $E_B$ protocols and
computed their yields. These in turn led to new lower bounds on
$Q_B$. The QECC that we studied, namely Cat code and Shor code,
exhibit different behaviors under this AQECC framework. For
example, for Shor code, it is beneficial to not insist on getting
no error in all measurements but instead carry out
error-correcting procedures after getting no error in the first
few measurements. Whereas for Cat code, one has to insist on
getting no error in all measurements. It is interesting to study
which of these two features is exhibited by other codes.

We also saw some connections with 2-EPP. Firstly, even though the
Leung-Shor method was introduced in \cite{LS1} as a 2-EPP, it is
in fact an $E_B$ protocol. Secondly, when the idea of modified
recurrence method is applied to Cat code, higher $E_B$ yields are
achieved.

Finally, one may want to ask whether the threshold fidelity in
section \ref{sec: threshold} goes down monotonically and if it
does, to what value it converges as n goes to infinity.

After the completion of this work, the conjectural relation
$Q_2>Q_B$ was proved \cite{N24}, and an emerging question is
whether the relation $Q_2>Q_B$ holds for all quantum channels
except when both capacities vanish. Also, can one show a
separation between $E_B$ and $Q_2$?

Acknowledgements: the author is grateful to Peter Shor for his
important insights during the course of this work, and would like
to thank Charles Bennett for his advice on the recent
developments.

\newpage 


\end{document}